\documentclass[prl,twocolumn]{revtex4}
\usepackage{graphicx}
\usepackage{setspace}

\begin{document}
	\bibliographystyle{apsrev}

	\title{Dynamics, efficiency and energy distribution of nonlinear plasmon-assisted generation of hot carriers}
	
	\author{O. Demichel, M. Petit, S. Viarbitskaya, R. Mejard,  F. de Fornel, E. Hertz,\\
		F. Billard,  A. Bouhelier, B. Cluzel}
	\vspace{5pt}
	
	\address{Laboratoire Interdisciplinaire Carnot de Bourgogne, 
		UMR 6303 CNRS-Universit\'e Bourgogne Franche-Comt\'e, 21078 Dijon, France}
	\begin{spacing}{1}
		
		\begin{abstract}
		We employ nonlinear autocorrelation measurements to investigate plasmon-assisted hot carrier dynamics generated in optical gold antennas. We demonstrate that surface plasmons enable a nonlinear formation of hot carriers, providing thus a unique lever to optimize the energy distribution and generation efficiency of the photo-excited charges.  The temporal response of the carriers' relaxation can be controlled within a range extending from 500~fs to 2.5~ps. By conducting a quantitative analysis of the dynamics, we determine the nonlinear absorption cross-section of individual optical antennas. As such, this work provides strong insights on the understanding of plasmon-induced hot carrier generation, especially in the view of applications where the time response plays a preponderant role. 
		\end{abstract}
		
		\maketitle
		Plasmonics--a general term associated with the collective charge oscillations induced by an electromagnetic excitation--aims at merging the ultrafast dynamics of electrons with the extreme localization of photons to open new avenues towards ultrafast-electronics \cite{Ozbay2006Science} or subwavelength light manipulation \cite{Barnes2003Nature}. As such, surface plasmons introduced a new paradigm in nanophotonics due to their ability to concentrate a far-field radiation into nanoscale domains. Plasmonics has already penetrated a large domain of applications ranging from single molecule sensing \cite{Xu1999PRL},  photothermal cancer therapy \cite{Loo2005NanoLett}, photovoltaics \cite{Atwater2010NatMat}, and more recently hot carrier harvesting~\cite{Sundararaman2014NatComm}

Upon excitation, energy can be transferred to carriers of metals, producing pairs of out-of-equilibrium electrons and holes that are referred to as hot carriers. Such hot carriers can for instance be produced by the absorption of a photon. Alternatively, plasmons may decay either radiatively by emitting a photon in the far field or through non-radiative electronic transitions which result in production of hot carriers. Recent theoretical \cite{Manjavacas2014Nano,Zhang2014JPhysChem,Sundararaman2014NatComm} and experimental \cite{Zheng2015NatComm} contributions demonstrated that a plasmon--assisted generation of hot carriers is much more efficient than a direct carrier photo-excitation. This make plasmonic nanostructures a very promising venue for developing the next generation of hot carrier technologies \cite{BrongersmaNatNano2015,Clavero2014NatPhot} with already demonstrated proof-of-concepts in photo-chemistry \cite{Takahashi2011APL}, photo-detection \cite{Stolz2014NanoLett,Sobhani2013NatComm} or photo-catalysis \cite{Mubeen2013NatNano} to name a few. However, these investigations are essentially focused at optimizing carrier harvesting through an energy level engineering, and are largely leaving aside the intrinsic dynamics of the carriers.  Yet, the timescales involved in hot carriers relaxation dictates the rate in which the energy is converted into heat \cite{Fan1992PRB,Groeneveld1995PRB,Sun1994PRB,DelFatti2000PRB,Baida2011PRL,Huang2014PNAS}  and play a fundamental role at determining the probability of transferring hot carriers into nearby acceptor levels or inducing chemical reactions. There are two other decisive parameters that should be optimized as well. They are the efficiency of the hot carrier generation and their energy distribution. Concerning the latter, only the decay of a single plasmon has been considered to date for a linear production of hot carriers \cite{Manjavacas2014Nano,Zhang2014JPhysChem,Sundararaman2014NatComm,Zheng2015NatComm,BrongersmaNatNano2015,Clavero2014NatPhot,Takahashi2011APL,Sobhani2013NatComm,Mubeen2013NatNano}. The energy of the nonequilibrium distribution is then intrinsically limited by the surface plasmon energy, e.g lower than 2.3 eV for gold-based plasmonics.  As for the efficiency, hot carriers formation is dictated by the electronic density of states (eDOS). For gold, the density of electron drastically increases at the onset the $d$-bands situated at a few eV below the Fermi level~\cite{Lasser1981PRB,Sundararaman2014NatComm}. Hence, a single linear plasmon excitation process does not allow for simultaneously controlling the hot carriers efficiency and energy distribution.  

In this letter, we overcome these limitations by tailoring the yield, the energy distribution and the dynamics through a nonlinear generation of photo-excited charges. Akin to a wide range of optical nonlinear effects observed in optical antennas,  we benefit from the near-field optical enhancement associated with the excitation of surface plasmons  \cite{Bouhelier2003APL,Imura2005JPhysChmeB,Kauranen2012NatPhot,Demichel2014OE}.
Using autocorrelation measurements of the nonlinear photoluminescence response of individual optical antenna~\cite{Biagioni2009PRB,Biagioni2012NanoLett,Jiang2013JPChemLett}, we probe the relaxation dynamics of hot carriers produced by the participation of three plasmons. The plasmon resonance and the optical pumping power positively contribute to increase the number of carriers. We find that these parameters are also affecting the dynamics, providing thus a leverage to extend the relaxation kinetics in the 500~fs to 2.5~ps range. We further quantify the efficiency of hot carrier generation by determining the nonlinear absorption cross section of resonant and off-resonant plasmonic antennas. 

Plasmonic gold nanorods antennas are fabricated by standard electron-beam lithography followed by metal deposition and lift-off. An array of gold nanorods with width of 55 nm and lengths varying from 90 nm to 800 nm as well as a plain thin gold  film are fabricated during the same process to ensure similar material quality and a common thickness of 35 nm. Localised surface plasmon resonances of such nanoantennas were previously reported in~\cite{Demichel2014OE}. The dynamics is probed by multiphoton photoluminescence (MPPL) auto-correlation measurements~\cite{Biagioni2009PRB,Biagioni2012NanoLett}. A Ti:Sapphire laser producings 120~fs pulses at 800~nm is focused into a diffraction-limited 300~nm spot by a high numerical aperture oil-immersion objective ($\times$60, NA 1.49). The nanorod antennas are positioned in the focal plane of the objective. The same objective collects the blue-shifted broad nonlinear photoluminescence, which is spectrally separated from the fundamental wavelength.  Autocorrelation measurements are achieved with a home-made Michelson interferometer. The incident laser beam is split into two arms before being recombined and focused on a single nanoantenna. One arm is motorized to control the inter-pulse delay with a femtosecond resolution. The dispersion induced along the optical path is carefully precompensated with a 4--f zero dispersion line~\cite{Weiner2011OptComm} to ensure Fourier transform limited chirp-free focused optical pulses.  Experiments are performed at room temperature ($\rm T_0 = 300$ K).

				\begin{figure}[t]
					\centering
					\includegraphics[width=0.45\textwidth]{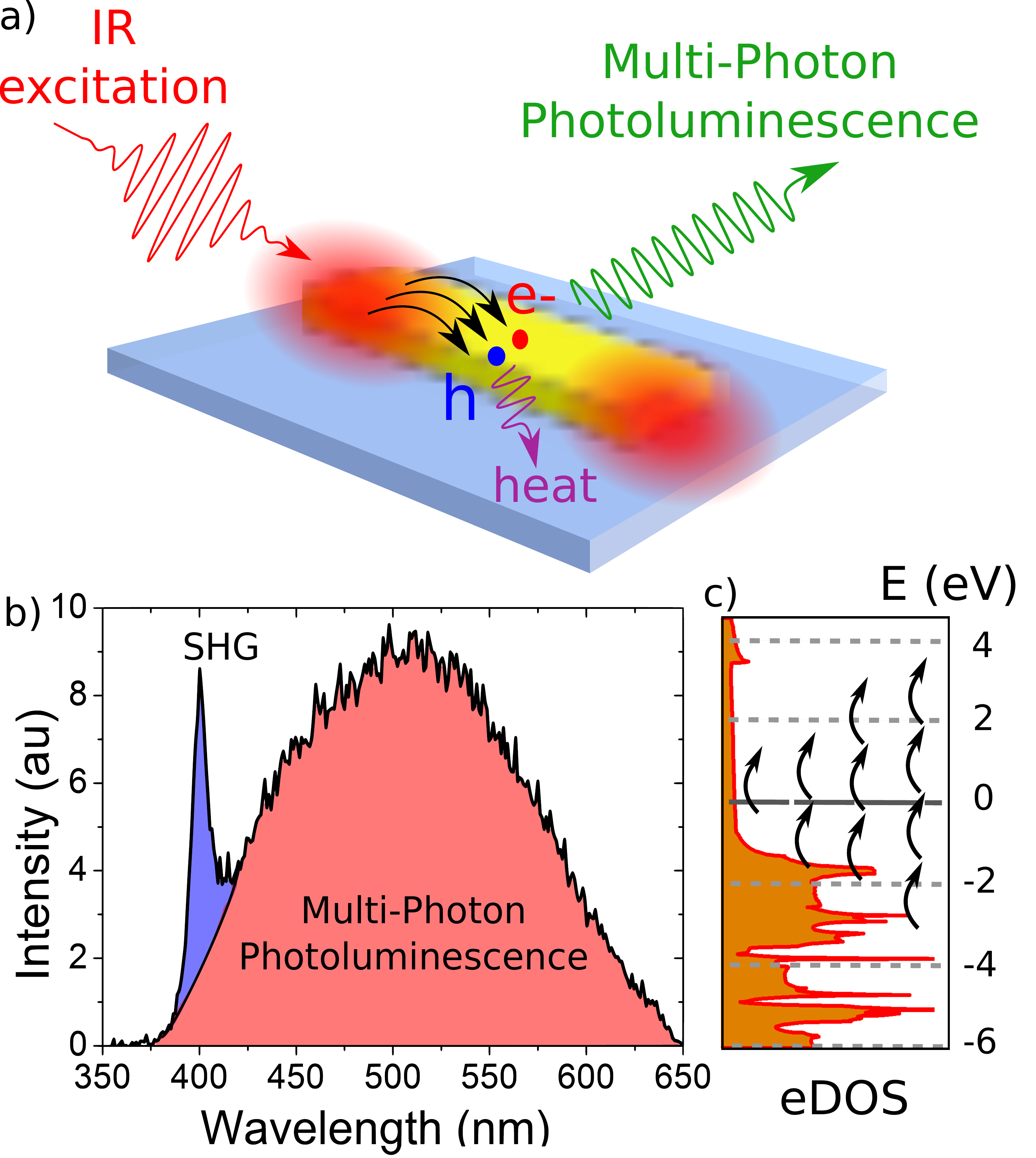}
					\caption{\label{manip} (a) Schematic view of the nonlinear plasmon-induced generation of hot carriers producing multiphoton photoluminescence (MPPL). (b) Typical nonlinear emission spectrum of a gold thin film showing the SHG at 400~nm and the broad band MPPL signal. (c) Representation of the gold's electronic DOS. The arrows indicate transitions assisted with respectively 1 to 4 plasmons.}
				\end{figure}

\begin{figure}[b!]
			\centering
			\includegraphics[width=0.45\textwidth]{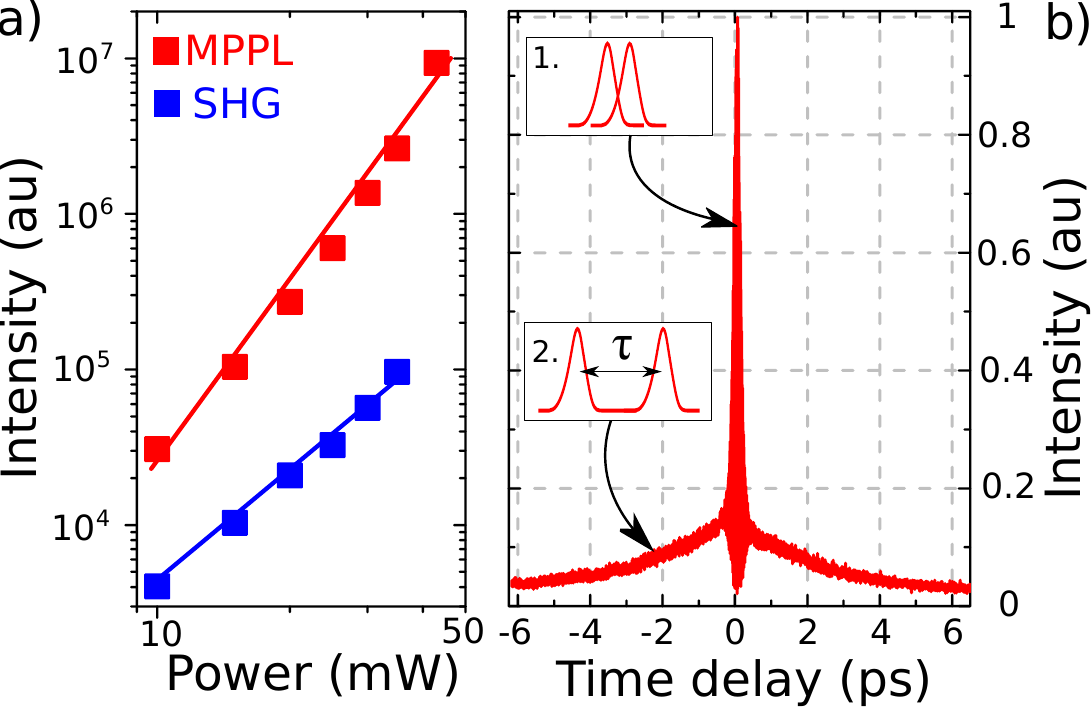}
			\caption{\label{auto-co} (a) Power dependence of SHG and MPPL intensity. (b) MPPL autocorrelation measurement from a gold nanorod. For short inter-pulse delay (inset 1.), the two pulses temporally overlap on the antenna resulting in an interference pattern. Longer delays reveal the dynamics of the MPPL intensity decay (inset 2.).}
		\end{figure}

Figure~\ref{manip}(a) sketches the plasmon-assisted MPPL mechanism in a gold nanorod and the decay of surface plasmons to hot carriers. MPPL is a nonlinear incoherent mechanism involving a sequential absorption mediated by a real intermediate state populated by hot carriers~\cite{Biagioni2012NanoLett}. The dynamic of the response is governed by the lifetime of the intermediate state and MPPL is thus a valuable tool to investigate plasmon-induced hot carrier generation efficiency and relaxation.  Figure~\ref{manip}(b) shows the typical nonlinear emission spectrum generated by the gold film. The spectrum is limited by the transparency window of the microscope objective and by the rejecting laser line filter. The nonlinear spectrum is composed of a broadband MPPL contribution together with a coherent second harmonic generation (SHG) at 400~nm. The electronic DOS for Au is depicted in Fig.~\ref{manip}(c). The number of vertical arrows for a given transition represents the order of the nonlinear process, and the length of the arrow corresponds to the laser energy (1.55~eV). The Fermi level defines the origin of the energy scale. The Fermi's golden rule says that the electronic transition rate is proportional to the joint electronic DOS of electrons and holes which appears to be minimal for the case of a single plasmon decay (one arrow). On the contrary, the decay of three or four plasmons is much more favoured by the high density of states available.
Then, a single plasmon excitation near the Fermi level and decaying in hot carriers is unfavoured compared to higher-order plasmon processes. In line with this argument, Fig.~\ref{auto-co}(a) shows a  MPPL nonlinearity order close to four, as already reported by Biagioni~\cite{Biagioni2012NanoLett}. The SHG nonlinearity is about 2, the non-integer value is due to the residual MPPL contribution overlapping the SHG peak (see spectra of Fig.~\ref{manip}(b)). The plasmon-induced hot carriers generated by the process can potentially reach energies as high as 4$\times$1.5~eV$\sim$6~eV.

Figure 2(b) illustrates a typical MPPL auto-correlation trace measured from a gold nanorod. In the region where pulses overlap (inset 1.), interferences are driving the MPPL intensity, hiding thus its ultrafast dynamics. In the following, we only consider delay longer than 450~fs when pulses are not temporally overlapping (inset 2.).  In this regime, the MPPL signal decreases exponentially with the inter-pulse delay. These wings indicate that the MPPL is a two-step mechanism that involves an intermediate state with a finite lifetime and the exponential decay probes this lifetime. In our hot carrier description of the MPPL mechanism, the first and the second pulse generate a hot carrier distribution that recombine by emitting a photon. However, the first pulse also populate the latter intermediate state with hot carriers which relax through carrier-phonon interactions. Then, for inter pulse delays shorter than the hot carrier relaxation time, plasmons generated by the second pulse can also decay by transferring their energy to these hot carriers produced by the first pulse. These re-excited carriers then recombine radiatively. The dynamics of these carrier-phonon interactions is probed by the MPPL autocorrelation traces~ \cite{Fan1992PRB,Sun1994PRB,Groeneveld1995PRB}.

\begin{figure}[t!]
					\centering

					\includegraphics[width=0.45\textwidth]{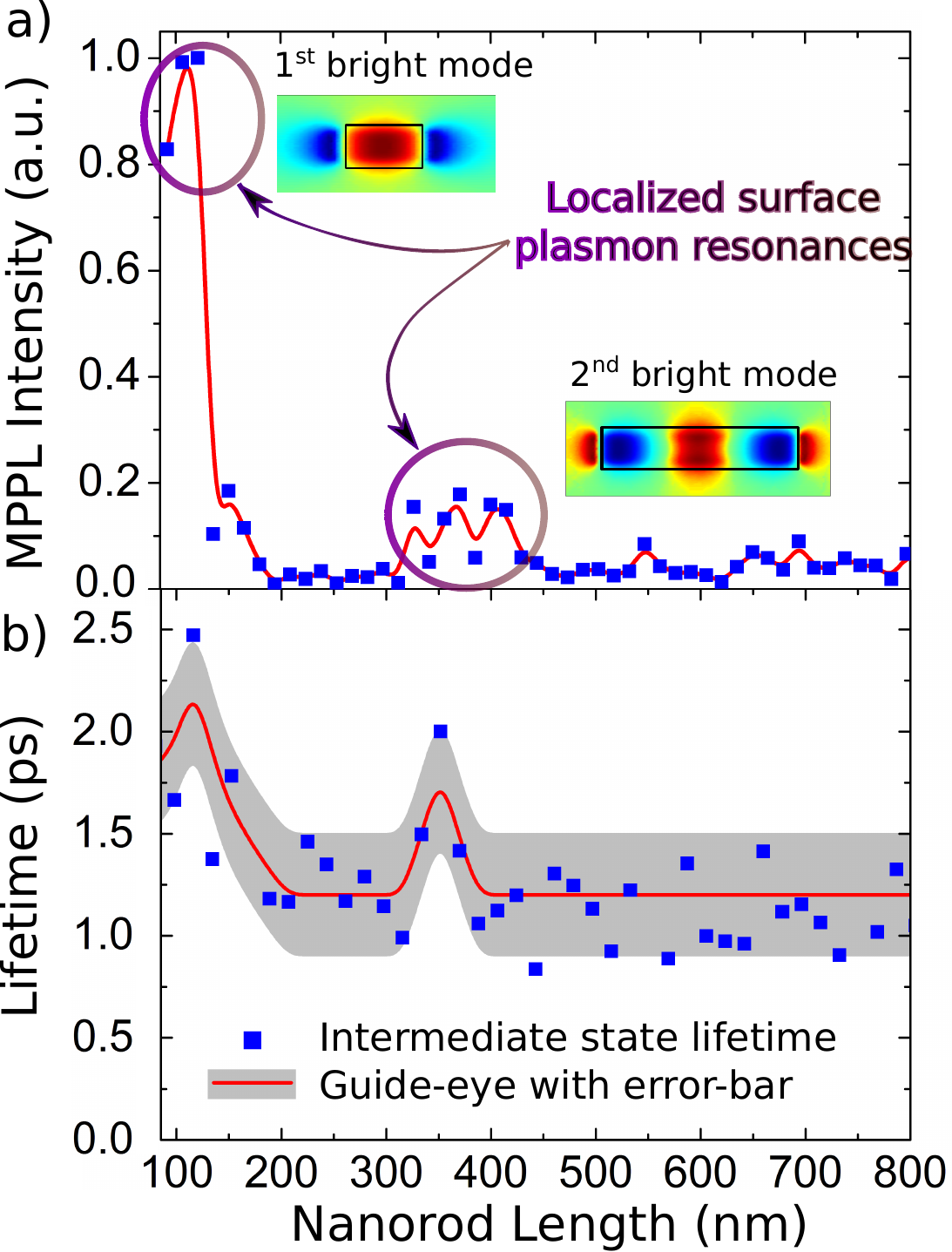}
					\caption{\label{lifetime-nanorods} (a) and (b) Evolution of the MPPL intensity and the intermediate state lifetime with the length of the nanorods, respectively. Insets: finite element numerical simulation of the electric field distribution parallel to the nanorod for $L$=100~nm and $L$=360~nm. }
				\end{figure}

We now turn to the fundamental questions addressed in this work, can one control plasmon-induced hot carrier dynamics? The  nonlinear process creating the hot carrier population is strongly affected by the local electromagnetic field enhancement. Hence, the generation efficiency can be independently optimized by shaping the geometry of the nanorod. We thus investigate the role played by plasmon resonances by probing the hot carrier lifetime for the fabricated array of gold nanoantennas with varying lengths. Figure~\ref{lifetime-nanorods}(a) and (b) show the MPPL intensity and hot carrier relaxation time as a function of the nanorod length $L$, respectively. There are two resonances highlighted by the MPPL signal corresponding to the first two bright modes~\cite{Demichel2014OE}. The simulated  distributions of the electric field component parallel to the nanorod long axis are reported in the insets.  A clear correlation exists between the MPPL intensity and the relaxation dynamics. The relaxation time is of 1.25 $\pm$ 0.25 ps for every off-resonant nanorods, whereas the dynamics for resonant antennas are enhanced by a factor two, reaching to 2.5 ps. Hence, by engineering the plasmon response of the antennas, the hot carrier lifetimes can be readily controlled.

To provide a deeper understanding, we provide in the following a description of hot carriers dynamics. Immediately after the decay of the plasmons into hot carriers, the photo-excited charges energy have a Dirac distribution that is then redistributed by carrier-carrier interactions following an internal thermalization process extensively discussed in Ref.~\cite{DelFatti2000PRB}. After this fast relaxation, experimentally hidden within the region where pulses interfere, the carrier distribution is  described by an electronic temperature $\rm T_{e,0}$~\cite{Fan1992PRB,Sun1994PRB,Groeneveld1995PRB}.  The carriers are then relaxing down to the Fermi level by transferring their energy to the lattice. Their energy distribution narrows and the associated electronic temperature ($\rm T_e$) decreases progressively~\cite{DelFatti2000PRB,Fan1992PRB,Sun1994PRB,Groeneveld1995PRB}. The process is generally described in the framework of the two temperature model defined by the rate equation \cite{DelFatti2000PRB}:
		\begin{equation}\rm
		C_e\frac{dT_e}{dt}=-\Gamma_{e,ph}(T_e-T_0),
		\end{equation}
		\noindent where $\rm\Gamma_{e,ph}$  is the carrier-phonon coupling constant. The initial electronic capacity is defined as $\rm C_e=c_oT_{e,0}$  in the weak perturbation approximation~\cite{DelFatti2000PRB} available here since the hot carrier dynamics always follows a single-exponential decays. $\rm c_o$ is the Sommerfeld constant.  According to that model, the hot carrier lifetime $\tau_e$ is written as \cite{DelFatti2000PRB,Baida2011PRL}:
		\begin{equation}\label{equation2}\rm
		\tau_e=\frac{c_0T_{e,0}}{\Gamma_{e,ph}}=\frac{c_0}{\Gamma_{e,ph}}\sqrt{T_0^2+\frac{2u_{abs}}{c_0}},
		\end{equation} 
where the initial temperature $T_{e,0}$ is expressed as a function of the room temperature ($\rm T_0$) and the energy absorbed per unit volume within a pulse ($\rm u_{abs}$). The apparition of $\rm T_0$ in the relation \ref{equation2} originates from the fact that before the impulse excitation, and also for low illuminating power, $T_{e,0}=\rm T_0$ in agreement with \cite{DelFatti2000PRB,Fan1992PRB,Sun1994PRB,Groeneveld1995PRB}.  The electronic temperature then increases with the number of generated carriers which produce stronger electron-electron interactions. This simple two-temperature model does not take into account any direct dependence on surface plasmons. Nevertheless, the term $\rm u_{abs}$ results from the product of the photon flux with the effective absorption cross-section of the nanorod, which is exalted at resonance~\cite{Pelton2008LPOR,Derom2012EPL}. In other words, the two-temperature model indirectly describes the plasmon dependence of the hot carrier dynamics.

				\begin{figure}[h!]
					\centering
					\includegraphics[width=0.45\textwidth]{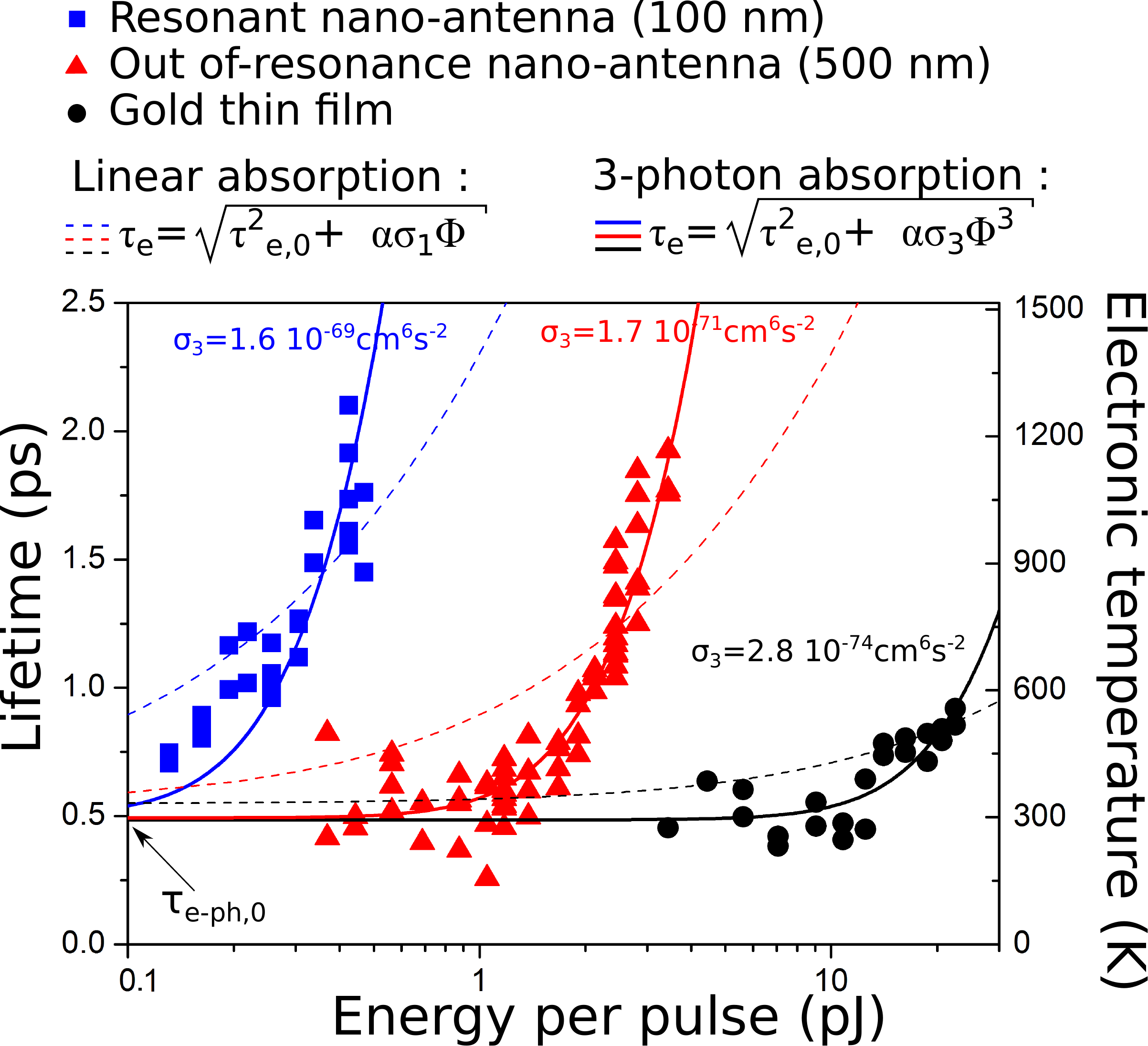}
					\caption{\label{model} Evolution of the intermediate lifetime $\tau_e$ with the pulse energy for a resonant antenna (blue squares), an out-of-resonance antenna (red triangles) and a planar gold thin film (dark circles). Solid and dashed coloured lines correspond to simulations based on the two temperature model, assuming that the intermediate state results respectively from a single or a three-photon absorption.}
				\end{figure}

According to Eq.~\ref{equation2}, $\rm\tau_e$ only depends on the absorbed energy per pulse ($\rm u_{abs}$) since other parameters are the lattice temperature and the properties of the material. Hence, the lifetime of the hot distribution must depend on the excitation power. We quantitatively compare in the following the power dependence for structures having different absorption cross-sections. Figure~\ref{model} display the dynamics for a 100 nm long resonant antenna (blue squares), a 500 nm long off-resonant nanorod (red triangles) and a planar thin film (black dots) as a function pulse energy.  Regardless of the type of structures, the lifetime is about 500~fs for low pulse energies. For gold, $\rm c_0$ = 66 J~m$^{-3}$~K$^{-2}$~\cite{Sun1993PRB} and $\rm\Gamma_{e,ph}$ = 4.10$^{16}$ W~m$^{-3}$~K$^{-1}$~\cite{Fan1992PRB}, thus in the low excitation regime, the thermalization time is given by $\rm\tau_{e,0} = \frac{c_0T_0}{\Gamma_{e,ph}}$ = 495~fs, in good agreement with our measurements. This value is consistent with the literature reporting hot carrier dynamics  in the 0.5--4~ps range, irrespective of the nonlinearity order of the MPPL process~\cite{Biagioni2009PRB,Biagioni2012NanoLett,Jiang2013JPChemLett,DelFatti2000PRB,Baida2011PRL}. For higher pulse energy, $\rm u_{abs}$ increases, and in agreement with Eq.~\ref{equation2}, we observe a concomitant raise of the lifetime. The evolution with the pulse energy underpins a dynamics governed by carrier-phonon interactions with a timescale directly related to the electron temperature after internal thermalization $\rm T_{e,0}$. The latter is readily inferred by feeding $\rm \tau_e$ in Eq.~\ref{equation2} and is reported on the right axis of Fig.~\ref{model}.   

Figure~\ref{model} shows that the dynamics of the thin film starts to be sensitive to the excitation power for a pulse energy two orders of magnitude higher than for a resonant nanorod. This is understood from the enhancement of the absorption cross-section ($\rm u_{abs}$) at resonance. The dashed lines plotted in Fig.~\ref{model} are the best fits to the experimental data assuming that the hot carriers are generated by a linear absorption. This description is clearly not reproducing the experimental trends. A better match to the data is obtained from a cubic dependence of $\rm u_{abs}$ on the pulse energy as shown by the solid lines of Fig.~\ref{model}.  The  absorption of three quanta is consistent with the order of the MPPL nonlinearity (Fig.~\ref{auto-co}(a)). We conclude that three plasmons are needed to populate the intermediate state with hot carriers after the first pulse, the second pulse providing only a fourth plasmon to the hot carrier distribution.
The absorbed energy per pulse delivered in the system writes as $\rm u_{abs} = 3\sigma_3h\nu\tau\Phi^3$, where $\rm \sigma_3$ refers to the three--photon absorption cross section, $\rm h\nu$ and $\rm \Phi$ are respectively the photon energy and flux, and $\rm\tau$ is the pulse duration. From the fits of Fig.~\ref{model}, we quantitatively evaluate $\rm \sigma_3$ for the two antennas and the thin film. We find $\rm \sigma_3=2.8 10^{-74}$ cm$^6$~s$^2$photon$^{-2}$ for the gold thin film, of  $\rm \sigma_3=1.7 10^{-71}$ cm$^6$~s$^2$photon$^{-2}$ for the non-resonant antenna and  $\rm \sigma_3=1.6 10^{-69}$ cm$^6$s$^2$photon$^{-2}$ for the resonant antenna. To our knowledge, these values are the first estimates for the three--photon absorption cross section of gold nanostructures. These numbers are orders of magnitude higher than those of single molecules and semiconductor nanoparticles which are respectively about 10$^{-80}$ cm$^6$~s$^2~$photon$^{-2}$~\cite{Liu2012JPCA}, and in the 10$^{-75}$ -- 10$^{-79}$ cm$^6~$s$^2$~photon$^{-2}$ range~\cite{Xing2011JPhysChemC}. 

In conclusion, we investigated the hot carrier dynamics generated in Au optical antennas through a plasmon-assisted mechanism. We found that the mediation of surface plasmons is essential to optimize the nonlinear generation of hot carriers. Specifically, through an increased optical absorption cross--section at the antenna's resonance, the efficiency of the hot-carrier generation increases by $100$-fold compared to off-resonances and by $10^5$ compared to thin films. The electronic temperature can be elevated above 1000~K, providing thus an enlarged energy distribution. We further demonstrate that the resonance retards the ultrafast relaxation of the photo-excited charges to the phonon  bath and can be as long as a few picoseconds.  Finally, we determined the nonlinear absorption cross--section of various plasmonic structures based on a quantitative analysis of the evolution of the relaxation dynamics with the incident pulse energy. Adjusting the different levers controlling the intrinsic properties of hot carriers (excitation efficiency, energy distribution and dynamics) and quantifying the relevant absorption cross--sections are of crucial importance  for advancing the next-generation of plasmon assisted hot carriers applications such as photo-chemistry, photo-detection or photo-catalysis. 

The research leading to these results has received fundings from the European Research Council under the European Community's Seventh Framework Program FP7/20072013 Grant Agreement no 306772 as well as the Agence Nationale de la Recherche (grants PLACORE ANR-13-BS10-0007 and CoConicS ANR-13-BS08-0013). This project has been performed in cooperation with the Labex ACTION program (contract ANR-11-LABX-0001-01).  O. D. and M. P. thank Thao D. for fruitful discussions and authors thanks V. Meunier for eDOS computations.


	\end{spacing}
\end{document}